\documentclass{eas}
\usepackage{graphicx}
\newcommand{\bu}{{\bf u}}

\newcommand{\be}{{\bf e}}
\newcommand{\bx}{{\bf x}}
\newcommand{\bk}{{\bf k}}

\begin{document}

\title{Double-diffusive convection} 
\author{Pascale Garaud}\address{Department of Applied Mathematics and Statistics, Baskin School of Engineering, University of California Santa Cruz, 1156 High Street, Santa Cruz, CA 95064}

\begin{abstract}
Much progress has recently been made in understanding and quantifying vertical mixing induced by double-diffusive instabilities
such as fingering convection (usually called thermohaline convection) and oscillatory double-diffusive convection (a process closely related to semiconvection). This was prompted in parts by advances in supercomputing, which allow us to run Direct Numerical Simulations of these processes at parameter values approaching those relevant in stellar interiors, and in parts by recent theoretical developments in oceanography where such instabilities also occur. In this paper I summarize these recent findings, and propose new mixing parametrizations for both processes that can easily be implemented in stellar evolution codes. 
\end{abstract}
\maketitle
\section{Introduction}
\label{sec:intro}

Double-diffusive instabilities commonly occur in any astrophysical fluid that is stable according to the Ledoux criterion, as long as the entropy and chemical stratifications have opposing contributions to the dynamical stability of the system. They drive weak forms of convection described below, and can cause substantial heat and compositional mixing in circumstances reviewed in this paper. Two cases can be distinguished. In {\it fingering convection}, entropy is stably stratified ($\nabla - \nabla_{\rm ad} < 0)$, but chemical composition is unstably stratified $(\nabla_\mu<0)$; it is  often referred to as {\it thermohaline} convection by analogy with the oceanographic context in which the instability was first discovered. In {\it oscillatory double-diffusive convection} (ODDC), entropy is unstably stratified ($\nabla - \nabla_{\rm ad} > 0)$, but chemical composition is stably stratified $(\nabla_\mu>0)$; it is related to semiconvection, but can occur even when the opacity is independent of composition. 

Fingering convection can naturally occur at late stages of stellar evolution, notably in giants, but also in Main Sequence stars that have been polluted by planetary infall (as first proposed by Vauclair, \cite{Vauclair04}), or by material transferred from a more evolved companion star. ODDC on the other hand is naturally found in stars in the vicinity of convective nuclear-burning regions, including high-mass core-burning Main Sequence or red clump stars, and shell-burning RGB and AGB stars. It is also thought to be common in the interior of giant planets that have been formed through the core-accretion scenario. 

\section{Linear theory and general considerations about vertical mixing}
\label{sec:linear}

Beyond competing entropy and compositional gradients, a necessary condition for double-diffusive instabilities to occur is 
$\tau \equiv \kappa_\mu/\kappa_T < 1$ ($\kappa_\mu$ and $\kappa_T$ are the microscopic compositional and thermal diffusivities). This is usually the case in astrophysical fluids, where $\tau$ is typically {\it much} smaller than one owing to the added contribution of photon and electron transport to the thermal diffusivity. 

\begin{figure}[t]
\includegraphics[width=\textwidth]{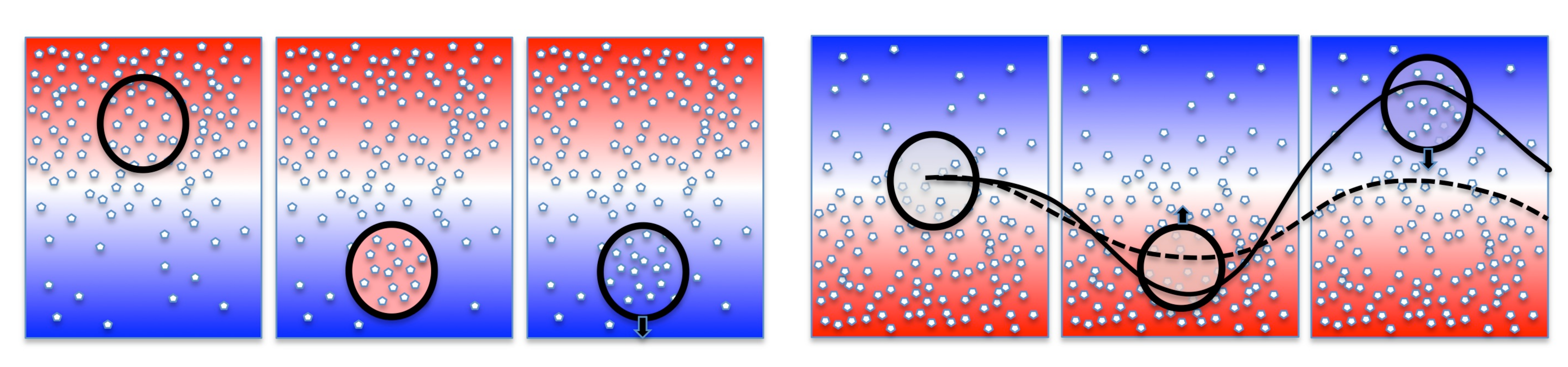} 
\caption{\small Illustration of the fingering (left) and oscillatory (right) double-diffusive instabilities (see main text for detail).}
\label{fig:instab}
\end{figure}

The somewhat counter-intuitive manner in which a high thermal diffusivity can be destabilizing is illustrated in Figure \ref{fig:instab}. In the fingering case, a small $\tau$ ensures that any small displaced fluid element rapidly adjusts to the ambient temperature of its new position, while retaining its original composition. An element displaced downward thus finds itself denser than the surrounding fluid and continues to sink; the opposite occurs for an element displaced upward.  In the case of ODDC, thermal diffusion can progressively amplify any internal gravity wave passing through, by heating a fluid element at the lowest position of its displacement and cooling it near the highest. In both cases, the efficient development of the instability is conditional on the fluid element being small enough for thermal diffusion to take place. Double-diffusive convection is therefore a process driven on very small scales, usually orders of magnitude smaller than a pressure scaleheight.

Consequently, a common way of studying fingering and ODD convection is by a {\it local} linear stability analysis, in which the background gradients of entropy (related to $\frac{dT_0}{dr} - \frac{dT_0^{\rm ad}}{dr} = T_0 (\nabla - \nabla_{\rm ad}) \frac{d \ln p_0}{dr}$) and composition $\frac{d\mu_0}{dr}$ are approximated as being constant (Baines \& Gill, \cite{BainesGill69}). The governing equations in the Boussinesq approximation are then:
\begin{eqnarray}
&& \frac{1}{{\rm Pr}}\left(\frac{\partial  \bu}{\partial t} +
  \bu \cdot \nabla  \bu\right) = -\nabla p
  + (T-\mu) \be_z +\nabla^2 \bu \mbox{   , }  \quad \nabla \cdot \bu = 0 \mbox{   , }  \nonumber  \\
&& \frac{\partial T}{\partial t} +  \bu \cdot \nabla
T \pm  w = \nabla^2 T \mbox{   , } \quad \frac{\partial \mu}{\partial t} +  \bu \cdot \nabla
\mu  \pm R_0^{-1}   w  = \tau \nabla^2 \mu \mbox{   , }
\label{eq:goveqs}
\end{eqnarray} 
where $\bu = (u,v,w)$, $p$, $T$  and $\mu$ are the non-dimensional velocity field, pressure, temperature and mean molecular weight perturbations of the fluid around the background state, Pr$=\nu/\kappa_T$ is the Prandtl number (and $\nu$ is the viscosity), and 
\begin{equation}
R_0 = \frac{\nabla - \nabla_{\rm ad}}{\frac{\phi}{\delta} \nabla_\mu} \mbox{   where  } \phi = \left( \frac{\partial \ln \rho}{\partial \ln \mu} \right)_{p, T}  \mbox{  and  } \delta = -\left( \frac{\partial \ln \rho}{\partial \ln T} \right)_{p, \mu} 
\end{equation}
 is called the {\it density ratio}. Here the unit lengthscale used is the typical horizontal scale $d$ of the basic instability, the unit time is $d^2/\kappa_T$, the unit temperature is $\Delta T_0 = |\frac{dT_0}{dr} - \frac{dT_0^{\rm ad}}{dr}|d$ and the unit compositional perturbation is $\frac{ \Delta T_0}{T_0}\frac{\delta}{\phi} \mu_0$. The $+$ sign in the temperature and composition equations should be used to model fingering convection, while the $-$ sign should be used to model ODDC. Mathematically speaking, this sign change is the only difference between the two processes.

Assuming perturbations have a spatio-temporal structure of the form $q(\bx,t)  = \hat q e^{i \bk \cdot \bx + \lambda t}$ where $q$ is either one of the dependent variables, $\bk$ is the wavenumber of the perturbation and $\lambda$ its growth rate (which could be complex), $\lambda$ satisfies a cubic equation: 
\begin{eqnarray}
&& \lambda^3 + k^2 (1 + {\rm Pr}+ \tau) \lambda^2  \\ 
&& +  \left[ k^4 (\tau +{\rm Pr} + \tau{\rm Pr}) \pm {\rm Pr} \frac{l^2}{k^2} (1-R_0^{-1}) \right] \lambda 
+ \left[ k^6 {\rm Pr} \tau \pm l^2 {\rm Pr} (\tau - R_0^{-1})  \right] = 0 \mbox{  , } \nonumber
\label{eq:cubic}
\end{eqnarray}
where $\pm$ again refers to $+$ for fingering convection and $-$ for ODDC, $k= |\bk|$ and $l$ is the norm of the horizontal component of $\bk$. $\lambda$ is real in the case of fingering convection but complex in the case of ODDC, as expected from the physical description of the mechanism driving the instability. The fastest growing mode in both cases is vertically invariant. Its growth rate $\lambda_{\rm fgm}$ and horizontal wavenumber $l_{\rm fgm}$ can be obtained by maximizing $Re(\lambda)$ over all possible $\bk$.  

Finally, setting $Re(\lambda_{\rm fgm})=0$ identifies marginal stability, and reveals the parameter range for double-diffusive instabilities to be:\begin{eqnarray}
&& 1 \le R_0 \le \tau^{-1} \mbox{   for fingering convection,}  \nonumber \\
&& \frac{{\rm Pr} + \tau}{{\rm Pr} + 1} \le R_0 \le 1 \mbox{   for ODDC.} 
\label{eq:criteria}
\end{eqnarray}
Note that $R_0 = 1$ in both cases corresponds to the Ledoux criterion.

While linear theory is useful to identify {\it when} double-diffusive convection occurs, nonlinear calculations are needed to determine how the latter saturates, and how much mixing it causes. Vertical mixing is often measured via non-dimensional vertical fluxes, called Nusselt numbers. The temperature and compositional Nusselt numbers are defined here as 
\begin{equation}
{\rm Nu}_T = 1 - \frac{F_T}{ \kappa_T  \left(\frac{dT_0}{dr} - \frac{dT_0^{\rm ad}}{dr}\right)} \mbox{   and   }{\rm Nu}_\mu = 1 - \frac{F_\mu}{ \kappa_\mu  \frac{d\mu_0}{dr} } \mbox{   , }
\end{equation}
where $F_T$ and $F_\mu$ are the dimensional temperature and compositional turbulent fluxes. To reconstruct the dimensional {\it total} fluxes of heat and composition ${\cal F}_T$ and ${\cal F}_\mu$, we have (Wood \etal, \cite{Woodal13})
\begin{equation} 
{\cal F}_T = - k_T \frac{d T_0}{dr} - ({\rm Nu}_T - 1)k_T \left(\frac{dT_0}{dr} - \frac{dT_0^{\rm ad}}{dr}\right) \mbox{   and   } {\cal F}_\mu = - {\rm Nu}_\mu \kappa_\mu  \frac{d\mu_0}{dr} \mbox{   , }
\label{eq:totalfluxes}
\end{equation}
where $k_T = \rho_0 c_p \kappa_T$ is the thermal conductivity, and $c_p$ is the specific heat at constant pressure. It is worth noting that ${\rm Nu}_\mu$ can also be interpreted as the ratio of the effective to microscopic compositional diffusivities.

Direct Numerical Simulations, which solve the fully nonlinear set of equations (\ref{eq:goveqs}) for given parameter values Pr, $\tau$ and $R_0$ from the onset of instability onward, can in principle be run to estimate the functions ${\rm Nu}_T(R_0;{\rm Pr},\tau)$ and Nu$_\mu(R_0;{\rm Pr},\tau)$. However, the actual nonlinear behavior of double-diffusive systems reveals a number of surprises, that must be adequately studied before a complete theory for mixing can be put forward.

\section{The emergence of large-scale structures}
\label{sec:large}

\subsection{Large-scale gravity waves and staircases}

\begin{figure}
\begin{center} \includegraphics[width=0.9\textwidth]{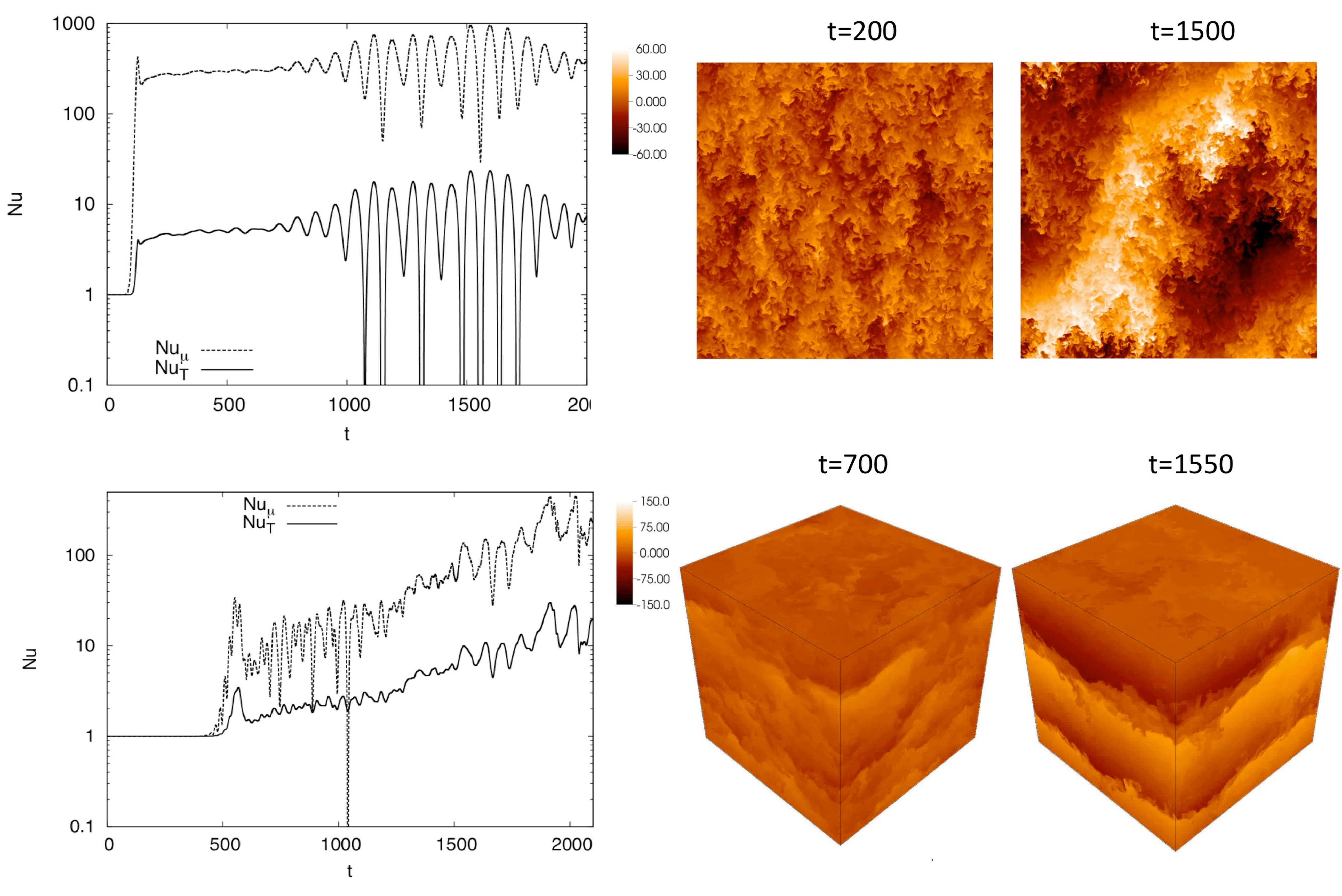} \end{center}
\caption{\small Top: Fingering simulation for Pr $=\tau=0.03$, $R_0 = 1.1$. The basic instability first saturates into a near-homogeneous state of turbulence, but later develops large-scale gravity waves. Bottom: ODDC simulation for Pr =$\tau=0.03$, $R_0 = 0.66$.  The basic instability first saturates into a near-homogeneous state, but later develops into a thermo-compositional staircase, whose steps gradually merge until only one is left. The mean Nusselt numbers increase somewhat in the presence of waves in the fingering case, and quite significantly when the staircase forms, and at each merger, in the ODDC case. }
\label{fig:large-scale}
\end{figure}

It has long been known in oceanography that double-diffusive convection has a tendency to drive the growth of structures on scales much larger than that of the basic instability (cf. Stern, \cite{Stern69}). This tendency was recently confirmed in the astrophysical context as well (Rosenblum \etal,  \cite{Rosenblumal11}; Brown \etal,  \cite{Brownal13}). These structures either take the form of  large-scale internal gravity waves or thermo-compositional staircases, as shown in Figure \ref{fig:large-scale}.

For density ratios close to one, fingering convection tends to excite large-scale gravity waves, through a process called the {\it collective instability} first discovered by Stern (\cite{Stern69}). These waves grow to significant amplitudes, and enhance mixing by fingering convection when they break. The same is true for ODDC, but the latter can sometimes also form thermo-compositional staircases excited by a process called the $\gamma-${\it instability} (Radko, \cite{Radko03}). The staircases spontaneously emerge from the homogeneously turbulent state, and appear as a stack of fully convective, well-mixed regions (the layers) separated by thin strongly stratified interfaces. The layers have a tendency to merge rather rapidly after they form. Vertical mixing increases significantly when layers form, and with each merger. 

For these reasons, quantifying transport by double-diffusive convection requires understanding not only how and at what amplitude the basic small-scale instabilities saturates, but also under which circumstances large-scale structures may emerge and how the latter affect mixing. 

\subsection{Mean-field theory}

Given their ubiquity in fingering and ODD convection, it is natural to seek a unified explanation for the emergence of large-scale structures that is applicable to both regimes. Mean-field hydrodynamics is a natural way to proceed, as it can capitalize on the separation of scales between the primary instability and the gravity waves or staircases. 

To understand how mean-field instabilities can be triggered, first note that the intensity of vertical mixing in double-diffusive convection is naturally smaller if the system is closer to being stable, and vice-versa. If a homogeneously turbulent state is spatially modulated by large-scale (but small amplitude) perturbations in temperature or chemical composition, then vertical mixing will be more efficient in regions where the {\it local} density ratio is closer to one, and smaller in regions where it is further from one. The spatial convergence or divergence of these turbulent fluxes can, under the right conditions, enhance the initial perturbations in a positive feedback loop, in which case a mean-field instability occurs. 

First discussed separately in oceanography, the collective and $\gamma$-instabilities were later discovered to be different unstable modes of the same mean-field equations by Traxler \etal~(\cite{Traxleral11a}), in the context of fingering convection. Their work has successfully been extended to explain the emergence of thermo-compositional layers in ODDC in astrophysical systems by Mirouh \etal~(\cite{Mirouhal12}).
A formal stability analysis of the mean-field equations shows that they are unstable to the $\gamma-$instability (the layering instability) whenever the {\it flux ratio} 
\begin{equation}
\gamma = \frac{R_0}{\tau} \frac{{\rm Nu}_T(R_0;{\rm Pr},\tau)}{{\rm Nu}_\mu(R_0;{\rm Pr},\tau)} 
\label{eq:gamma}
\end{equation}
is a {\it decreasing} function of $R_0$ (Radko, \cite{Radko03}). 
Similarly, a necessary condition for the collective instability was given by Stern \etal~(\cite{Sternal01}), who argued\footnote{It is worth noting that whether his criterion, which was derived in the context of physical oceanography, applies {\it as is} in the astrophysical regime has not yet been verified.} that large-scale gravity waves can develop whenever 
\begin{equation}
A = \frac{({\rm Nu}_T-1)(\gamma_{\rm turb}^{-1}-1)}{{\rm Pr}(1-R_0^{-1})} > 1 \mbox{   , }
\label{eq:A}
\end{equation}
where $\gamma_{\rm turb} = (R_0/\tau) ({\rm Nu}_T-1)/({\rm Nu}_\mu-1)$. This criterion is often much less restrictive than the one for the development of the $\gamma-$instability. Note that ${\rm Nu}_T$ and ${\rm Nu}_\mu$ in (\ref{eq:gamma}) and (\ref{eq:A}) are the Nusselt numbers associated with the small-scale turbulence present {\it before} any large-scale structure has emerged.  

\section{Fingering (thermohaline) convection}

\subsection{Saturation of the primary instability}

Traxler \etal~(\cite{Traxleral11b}) were the first to run a systematic sweep of parameter space to study fingering convection in astrophysics, and to measure ${\rm Nu}_T(R_0;{\rm Pr},\tau)$ and ${\rm Nu}_\mu(R_0;{\rm Pr},\tau)$ in 3D numerical experiments. However, they were not able to achieve very low values of Pr and $\tau$. Brown \etal~(\cite{Brownal13}) later presented new simulations with Pr and $\tau$ as low as $10^{-2}$, but this is still orders of magnitude larger than in stellar interiors, where Pr and $\tau$ typically range from $10^{-8}$ to $10^{-3}$.

To bridge the gap between numerical experiments and stellar conditions, Brown \etal~(\cite{Brownal13}) derived a compelling semi-analytical prescription for transport by small-scale fingering convection, that reproduces their numerical results and can be extrapolated to much lower Pr and $\tau$. Their model attributes the saturation of the fingering instability to the development of shearing instabilities between adjacent fingers (see also Denissenkov, \cite{Denissenkov10} and Radko \& Smith, \cite{RadkoSmith12}).

For a given set of governing parameters Pr, $\tau$ and $R_0$, the growth rate and horizontal wavenumber of the fastest-growing fingers can be calculated from linear theory (see Section \ref{sec:linear}). Meanwhile, the growth rate $\sigma$ of shearing instabilities developing between neighboring fingers is proportional to the velocity of the fluid within the finger times its wavenumber (a result that naturally emerges from dimensional analysis, but can also be shown formally using Floquet theory). Stating that shearing instabilities can disrupt the continued growth of fingers requires $\sigma$ and $\lambda_{\rm fgm}$ to be of the same order. This sets the velocity within the finger to be $W  = C \lambda_{\rm fgm}/l_{\rm fgm}$ where $C$ is a universal constant of order one. Meanwhile, linear stability theory also relates the temperature and compositional fluctuations $T$ and $\mu$ within a finger to $W$. The turbulent fluxes can thus be estimated {\it only using linear theory}:
\begin{equation}
{\rm Nu}_T =1 + \frac{1}{ l_{\rm fgm}^2 } \frac{C^2\lambda_{\rm fgm}^2}{\lambda_{\rm fgm} +  l_{\rm fgm}^2} \mbox{  ,  } 
{\rm Nu}_\mu = 1 + \frac{1}{\tau  l_{\rm fgm}^2 } \frac{C^2\lambda_{\rm fgm}^2}{\lambda_{\rm fgm} + \tau l_{\rm fgm}^2} \mbox{   . }
\label{eq:Nusfinger}
\end{equation}
Comparison of these formula with the data helps calibrate $C$. Brown \etal~(\cite{Brownal13}) found that using $C = 7$ can very satisfactorily reproduce most of their data within a factor of order one or better, except when Pr $< \tau$ (which is rarely the case in stellar interiors anyway). 

Equation (\ref{eq:Nusfinger}) implies that for low Pr and $\tau$, turbulent heat transport is negligible, while turbulent compositional transport is significant only when $R_0$ is close to one.  However, the values of ${\rm Nu}_\mu$ obtained by Brown \etal~(\cite{Brownal13}) are still not large enough to account for the mixing rates required by Charbonnel \& Zahn (\cite{CharbonnelZahn07}) to explain surface abundances in giants. Such large values of $D_\mu$ might on the other hand be achieved if mean-field instabilities take place. 

\subsection{Mean-field instabilities in fingering convection}

As discussed in Section \ref{sec:large}, one simply needs to estimate $\gamma$ and $A$ in order to determine in which parameter regime mean-field instabilities can occur. Using (\ref{eq:gamma}) and (\ref{eq:Nusfinger}) it can be shown that $\gamma$ is always an {\it increasing} function of $R_0$ at low Pr and $\tau$. This implies that fingering convection is stable to the $\gamma-$instability, and therefore not likely to transition {\it spontaneously} to a state of layered convection in astrophysics. The simulations of Brown \etal~(\cite{Brownal13}) generally confirm this statement, except in a few exceptional cases discussed below.

By contrast, fingering convection does appear to be prone to the collective instability (as shown in Figure \ref{fig:large-scale}) for sufficiently low $R_0$. By calculating $A$ for a typical stellar fluid with Pr $\sim 10^{-6}$ and $\tau \sim 10^{-7}$, we find for instance that gravity waves should emerge when $R_0 < 100$ or so. In this regime, we expect transport to be somewhat larger than for small-scale fingering convection alone, although probably not by more than a factor of 10 (see Figure \ref{fig:large-scale}). Nevertheless, a first-principles theory for the vertical mixing rate in fingering convection, in the presence of internal gravity waves, remains to be derived.   

Finally, as first hypothesized by Stern (\cite{Stern69}) and found in preliminary work by Brown \etal~(\cite{Brownal13}), it is possible that these large-scale gravity waves could break on a global scale and mechanically drive the formation of layers. If this is indeed confirmed, transport could be much larger than estimated in (\ref{eq:Nusfinger}) in the region of parameter space for which $A<1$. This, however, remains to be confirmed. 

\subsection{A plausible 1D mixing prescription for fingering convection}

Until we gain a better understanding of the various effects of gravity waves described above, (\ref{eq:Nusfinger}) is our best current estimate for transport by fingering convection in astrophysical objects. An example of the numerical implementation of the model by Brown \etal~(\cite{Brownal13}) is now available in MESA, and consists of the following steps. (1) To estimate the local properties of the star, and calculate all governing parameters/diffusivities. (2) To estimate the properties of the fastest-growing fingering modes using linear theory (see Section \ref{sec:linear}) and (3) To apply (\ref{eq:Nusfinger}) to calculate ${\rm Nu}_\mu$, and then (\ref{eq:totalfluxes}) to calculate ${\cal F}_\mu$. Turbulent heat transport is negligible, so ${\cal F}_T = - k_T dT_0/dr$.

\section{Oscillatory double-diffusive convection (semiconvection)}

\subsection{Saturation of the primary instability}

3D numerical simulations of ODDC were first presented by Rosenblum \etal~(\cite{Rosenblumal11}) and Mirouh \etal~(\cite{Mirouhal12}). Both explored parameter space to measure, as in the case of fingering convection, the functions ${\rm Nu}_T(R_0;{\rm Pr},\tau)$ and ${\rm Nu}_\mu(R_0;{\rm Pr},\tau)$ after saturation of the basic ODD instability. The values of Pr and $\tau$ achieved, however, were not very low, and models are needed once again to extrapolate these results to parameters relevant for stellar interiors. Mirouh \etal~(\cite{Mirouhal12}) proposed an empirical formula for Nu$_T(R_0;{\rm Pr},\tau)$ (and Nu$_\mu(R_0;{\rm Pr},\tau)$, via $\gamma$), whose parameters were fitted to the experimental data. However, a theory based on first principles is more desirable.

We have recently succeeded in applying a very similar method to the one used by Brown \etal~(\cite{Brownal13}) to model transport by small-scale ODDC. As described in Moll \etal~(\cite{Mollal14}), a simple approximate estimate for temperature and compositional Nusselt numbers can also be derived from the linear theory for the fastest-growing mode (see Section \ref{sec:linear}), this time in the form of 
\begin{equation} 
{\rm Nu}_T = 1 + \frac{\lambda_R \tilde{\lambda} }{l_{\rm fgm}^2} \frac{ \lambda_R + l_{\rm fgm}^2 }{( \lambda_R + l_{\rm fgm}^2)^2  + \lambda_I^2}  \mbox{   , } {\rm Nu}_\mu = 1 + \frac{\lambda_R \tilde{\lambda}  }{\tau l_{\rm fgm}^2} \frac{\lambda_R + \tau l_{\rm fgm}^2 }{( \lambda_R + \tau l_{\rm fgm}^2)^2  + \lambda_I^2} \mbox{   , }
\label{eq:NusODDC}
\end{equation}
where $\lambda_R= Re(\lambda_{\rm fgm})$, $\lambda_I=Im(\lambda_{\rm fgm})$ and $\tilde{\lambda} = \sqrt{K_1^2 \lambda_R^2 + K_2^2 \lambda_I^2}$. The constants $K_1$ and $K_2$ must again be fitted to the existing data; preliminary results suggest that $K_1 \simeq 50$ and $K_2 \simeq 7$.

By contrast with fingering convection, ODDC is subject to both $\gamma-$ and collective instabilities. Both modify the vertical heat and compositional fluxes quite significantly so (\ref{eq:NusODDC}) should {\it not} be used {\it as is} to model mixing by ODDC. It is used, on the other hand, to determine when mean-field instabilities occur.

\subsection{Layered convection}

The region of parameter space unstable to layering can again be determined by calculating $\gamma$ (using \ref{eq:NusODDC} this time), and checking when $d\gamma/dR_0 < 0$. Moll \etal~(\cite{Mollal14}) (see also Mirouh \etal, \cite{Mirouhal12}) showed that layering is always possible for Pr and $\tau$ below one, provided $R_0 \in [R_c({\rm Pr},\tau), 1]$. The critical value $R_c({\rm Pr},\tau)$ is fairly close to zero in the parameter regime appropriate for stellar interiors, as $R_c \propto {\rm Pr}^{1/2}$ in the limit $\tau \sim {\rm Pr} \rightarrow 0$. This implies that the region of parameter space unstable to layer formation spans nearly the entire ODDC range.

As discussed in Section \ref{sec:large}, the vertical heat and compositional fluxes increase dramatically when a staircase first emerges, and then again at each layer merger. This was studied by Wood \etal~(\cite{Woodal13}), who ran and analyzed simulations for $R_0 \in [R_c,1]$, and argued that their results are consistent with the following empirical transport laws for layered convection:
\begin{equation}
{\rm Nu}_T = 1 + g_T(R_0,\tau) ({\rm Ra} {\rm Pr})^{1/3} \mbox{   and   } {\rm Nu}_\mu = 1 + g_\mu(R_0,\tau) {\rm Ra}^{0.37}  {\rm Pr}^{1/4}  \tau^{-1} \mbox{   , }
\label{eq:Nuslayers}
\end{equation}
where $g_T$ and $g_\mu$ are slowly varying functions of $R_0$ and $\tau$, and where the Rayleigh number is ${\rm Ra} = \left| \frac{\delta }{\rho_0} \frac{d p_0}{dr} \left(\frac{d \ln T_0}{dr} - \frac{d \ln T_{\rm ad}}{dr} \right) \frac{H^4}{ \kappa_T \nu} \right| $, where $H$ is the mean step height in the staircase. For numerically achievable Pr and $\tau$, Wood \etal~(\cite{Woodal13}) estimated that $g_T \simeq 0.1$ and $g_\mu \simeq 0.03$.

Equation (\ref{eq:Nuslayers}) has two important consequences. The first is that turbulent heat transport can be significant in layered convection, provided $H$ is large enough. Secondly, both Nusselt numbers are (roughly) proportional to $H^{4/3}$, but nothing so far has enabled us to determine what $H$ may be in stellar interiors. Indeed, in {\it all} existing simulations of layered ODDC to date, layers were seen to merge fairly rapidly until a single one was left. Whether staircases in stellar interiors evolve in the same way, or eventually reach a stationary state with a layer height smaller than the thickness of the unstable region itself, is difficult to determine without further modeling, and remains the subject of current investigations.

\subsection{Large-scale gravity waves}

ODD systems which do not transition into staircases ($R_0 < R_c$) also usually evolve further with time after saturation of the basic instability, with the small-scale wave-turbulence gradually giving way to larger-scale gravity waves. Whether the latter are always excited by the collective instability, or could be promoted by other types of nonlinear interactions between modes that transfer energy to larger scales, remains to be determined. In all cases, these large-scale waves have significant amplitudes and regularly break. This enhances transport, as it did in the case of fingering convection. Moll \etal~(\cite{Mollal14}) found that the resulting Nusselt numbers are between 1.2 and 2 across most of the unstable range, regardless of Pr or $\tau$. 
These results are still quite preliminary, however, and their dependence on the domain size (which sets the scale of the longest waves) remains to be determined.

\subsection{A plausible 1D mixing prescription for ODDC}

Based on the results obtained so far, and summarized above, a plausible mixing prescription for ODDC can be obtained
by applying the following steps. (1) To estimate the local properties of the star, and calculate all governing parameters/diffusivities. (2) To estimate the properties of the fastest-growing modes using linear theory (see Section \ref{sec:linear}) (3) To determine whether layers are expected to form or not by calculating $\gamma$ (using \ref{eq:NusODDC}) for neighboring values of $R_0$, and evaluating $d\gamma/dR_0$. (4) If the system is layered, then {\it assume a layer height} (for instance, some small fraction of a pressure scaleheight), and calculate the heat and compositional fluxes using (\ref{eq:totalfluxes}) with (\ref{eq:Nuslayers}). If the system is not expected to form layers, then calculate these fluxes using (\ref{eq:totalfluxes}) and ${\rm Nu}_T \sim {\rm Nu_\mu} \sim 1.2-2$ instead. The unknown layer height is the only remaining free parameter of this model, and will hopefully be constrained in the future by comparison of the model predictions with asteroseismic results.

\begin{figure}
\begin{center} \includegraphics[width=0.92\textwidth]{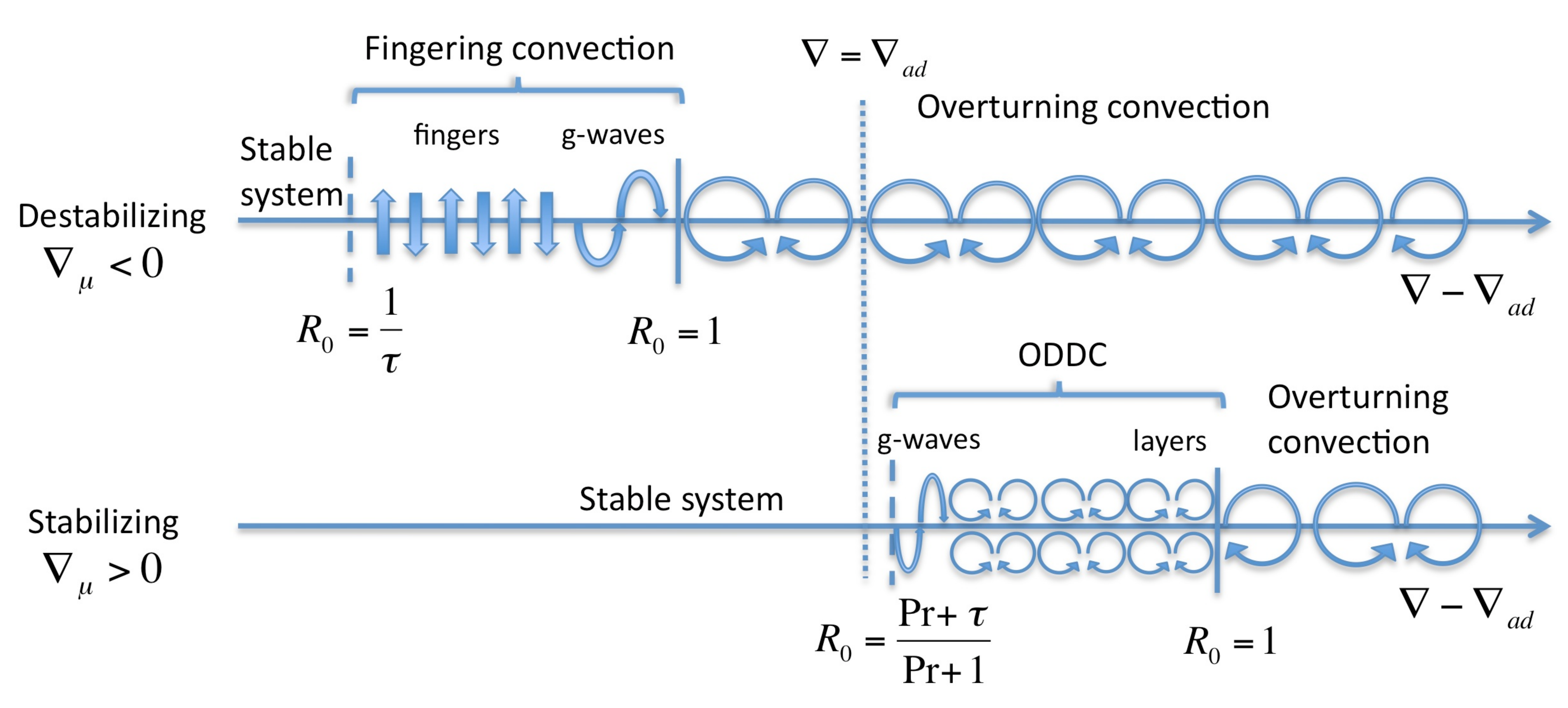} \end{center}
\caption{\small Illustration of the various regimes of double-diffusive convection in astrophysics discussed in this paper, as we understand them today.}
\label{fig:regimes}
\end{figure}


\begin{thebibliography}{99}

\bibitem[1969]{BainesGill69} Baines, P. \& Gill, A. 1969, JFM, 37, 289

\bibitem[2013]{Brownal13} Brown, J.~M., Garaud, P., \& Stellmach, S. 2013, ApJ, 768, 34

\bibitem[2007]{CharbonnelZahn07} Charbonnel, C. \& Zahn, J. 2007, A\&A, 467, L15

\bibitem[2010]{Denissenkov10} {Denissenkov}, P.~A. 2010, ApJ, 723, 563

\bibitem[2012]{Mirouhal12} Mirouh, G., Garaud, P., Stellmach, S., Traxler, A., \& Wood, T. 2012, ApJ, 750, 61

\bibitem[2014]{Mollal14} Moll, R., Garaud, P., Stellmach, S. 2014, in prep.

\bibitem[2003]{Radko03} Radko, T. 2003, JFM, 497, 365

\bibitem[2012]{RadkoSmith12} {Radko}, T. \& {Smith}, D.~P. 2012, JFM, 692, 5

\bibitem[2011]{Rosenblumal11} {Rosenblum}, E., {Garaud}, P., {Traxler}, A., \& {Stellmach}, S., 2011,ApJ, 731, 66

\bibitem[1969]{Stern69} Stern, M. 1969, JFM, 35, 209

\bibitem[2001]{Sternal01} Stern, M., Radko, T., \& Simeonov, J. 2001, J. Mar. Res., 59, 355

\bibitem[2011a]{Traxleral11a} {Traxler}, A., {Stellmach}, S., {Garaud}, P., {Radko}, T., \& {Brummell}, N. 2011a, JFM, 677, 530

\bibitem[2011b]{Traxleral11b} Traxler, A., Garaud, P., \& Stellmach, S. 2011b, ApJL, 728, L29

\bibitem[2004]{Vauclair04} Vauclair, S. 2004, ApJ, 605, 874

\bibitem[2013]{Woodal13} Wood, T.~S., Garaud, P., \& Stellmach, S. 2013, ApJ, 768, 157

\end{thebibliography}

\end{document}